\documentclass[12pt]{article}
\usepackage{babbage}
\usepackage{graphicx}
\begin{document}

\title{Coloured conical emission via second-harmonic generation}

\author{S. Trillo,\cite{byline1} C. Conti}
\address{
Istituto Nazionale di Fisica della Materia (INFM)-RM3,\\
Via della Vasca Navale 84, 00146 Roma, Italy}

\author{P. Di Trapani, O. Jedrkiewicz, J. Trull \cite{byline2}}
\address{INFM and Dept. of Chemical, Physical and  Mathemetical Sciences,\\
University of Insubria, Via Valleggio 11, 22100 Como, Italy}

\author{G. Valiulis}
\address{Dept. of Quantum Electronics, Vilnius University,\\
Building 3 Sauletekio Avenue 9, bldg. 3, LT-2040, Vilnius, Lithuania}

\author{G. Bellanca}
\address{Dept. of Engineering, University of Ferrara, Via
Saragat 1, 44100 Ferrara, Italy}

\newpage
\begin{abstract}
We predict that the combination of space and time modulational
instabilities occurring via parametric wave-mixing in quadratic
media leads to coloured conical emission. This phenomenon should
be observed under conditions usually employed in second-harmonic
generation experiments.
\end{abstract}

Modulational instability (MI) stands for the exponential growth of
periodic, long-wave\-length  perturbations occurring either in time
(i.e., temporal break-up \cite{tai86}),  or space (filamentation,
e.g. Ref. [2]), due to interplay of Kerr nonlinearity and
group-velocity dispersion (GVD) or diffraction, respectively. MIs
occur also in quadratic media, in the form of parametric
instabilities of two-color pump beams, encompassing both the cases
of propagation invariant solutions (i.e., eigenmodes
\cite{trillo95,schiek01}), and periodic pump evolutions
\cite{fuerst97,tw97} as in, e.g. unseeded second-harmonic
generation (SHG). Filamentation has been observed both in unseeded
and seeded SHG geometries, in 1+2 (bulk) and 1+1 (planar
waveguides) dimensions \cite{schiek01,fuerst97,fang00}, while
temporal MI have not been observed yet. On this ground, one might
naturally wonder about the coupling of spatial and temporal
degrees of freedom, as occurs, e.g., in self-focusing of short
pulses. In this letter, our purpose is to show that, in the
presence of both dispersion and diffraction, spatio-temporal MI in
quadratic bulk samples cannot be simply seen as the mere
superposition of filamentation and temporal break-up. Rather, it
leads to a new phenomenon, namely coloured conical emission (CCE),
where rings of different colours are expected to grow
exponentially on top of both pump far-fields. As we show CCE via
SHG has specific features which make it different from that
earlier predicted for Kerr media \cite{liou92,luther94,fonseca99}.
It differs also from other space-time phenomena characteristic of
SHG, such as snake instabilities \cite{derossi97}, where the
spatial breakup of SHG solitons associated with temporal MI does
not lead to substantial modifications of the beam angular spectra,
which remain confined nearly on-axis.
\newline\indent
Although the phenomenon of CCE implies
component which travel off-axis, the angles usually involved are
rather small and it is reasonable to use the following paraxial dimensionless model
for SHG in noncritical configuration (i.e., no spatial walk-off)
\begin{eqnarray}
&i&\frac{\partial u_1}{\partial z} +
\frac{\sigma_1}{2}\nabla_{\perp}^2 u_1
-\frac{\beta_1}{2} \frac{\partial^2 u_1}{\partial t^2}
+ u_2 u_1^* =  0,\nonumber \\
\label{SHGdimless} \\
&i&\frac{\partial u_2}{\partial z} +
\frac{\sigma_2}{2} \nabla_{\perp}^2 u_2 +
i \delta \frac{\partial u_2}{\partial t}
-\frac{\beta_2}{2} \frac{\partial^2 u_2}{\partial t^2}  +
\delta k u_2 + \frac{u_1^2}{2} = 0,\nonumber
\end{eqnarray}
\noindent Here, the (cw plane-wave) fields are conveniently
normalized in terms of their total intensity $I_t$ to yield
$|u_2|^2+|u_1|^2/2=1$. Accordingly the real-world distance $Z=z
Z_{nl}$ is fixed by the nonlinear length scale $Z_{nl}=(\chi
I_t)^{-1}$, where $\chi=k_0 (2 \eta_0/n_1^2 n_2)^{1/2} d_{eff}$
\cite{trillo95}. The normalized retarded time $t=(T-Z/V_{g1})/T_0$
and the Laplacian $\nabla_{\perp}^2=\partial_x^2+\partial_y^2$,
where $(x,y)=(X,Y)/R_0$, are given in terms of temporal and
transverse scales, $T_0=(|k_1''| Z_{nl})^{1/2}$ and
$R_0=(Z_{nl}/k_1)^{1/2}$, respectively. Moreover
$\sigma_m=k_1/k_m$, $m=1,2$ (we set $\sigma_2 = \frac{1}{2}$),
$\beta_{m}=k_m''/|k_1''|$,  $k''_m=d^2k/d\omega^2|_{m\omega}$ are
GVDs, $\delta k=(k_2-2k_1) Z_{nl}$ is the phase mismatch, and
$\delta=Z_{nl}/Z_w$ stands for group-velocity mismatch (GVM),
$Z_w=T_0 (1/V_{g2}-1/V_{g1})^{-1}$ being the walk-off length
($V_{gm}=dk/d\omega|^{-1}_{m\omega}$).

Let us consider propagation-invariant solutions of Eqs.~(\ref{SHGdimless}),
i.e. cw plane-wave eigenmodes $u_m=u_{m0}^{\pm} \exp(i m \beta z)$, $m=1,2$,
where $u_{20}^{\pm}=\beta=\left[ \delta k \pm (\delta k^2 + 12)^{1/2}
\right]/6$, $u_{10}^{\pm 2}=2[1-|u_{20}^{\pm}|^2]$, the
upper and lower sign standing for in-phase
($\phi={\rm arg}(u_2-2u_1)=0$, existing for $-\infty<\delta k \le 2$)
and out-of-phase ($\phi=\pi$, $\infty >\delta k \ge -2$ )
modes, respectively \cite{trillo95}.
We study the stability of these waves
by inserting in Eqs.~(\ref{SHGdimless}) the fields
\begin{equation} \label{ansatz}
u_m(r,t,z)=[u_{m0} + a_m(r,t,z)] \exp(i m \beta z), m=1,2,
\end{equation}
\noindent where $u_{m0}=u_{m0}^{\pm}$,
and we consider a radially symmetric perturbation:
\begin{equation}
a_m(r,t,z)=f_m^+(r,t) \exp(\lambda z)
+  f_m^-(r,t)^{\ast} \exp(\lambda^{\ast} z).
\end{equation}
After linearization, we obtain a system of four
partial differential equations in the unknowns $f_{1,2}^{\pm}(r,t)$.
Without loss of generality, we introduce the Fourier-Bessel transform of $f_m^{\pm}$,
$F_m^{\pm} (K,\Omega) = {\cal F} \left[  f_m^{\pm} \left( r,t \right) \right]$,
which permits to express the perturbation amplitudes
through the inverse transform
$f_m^{\pm} = {\cal F}^{-1} \left[  F_m^{\pm} \right]$,
as a superposition of conical $J_0$
Bessel waves \cite{durnin87} of transverse wavevector
$K=[K_x^2+K_y^2]^{1/2}$ and frequency $\Omega$, as
\begin{equation} \label{transform}
f_m^{\pm} \left( r,t \right) = \int_{-\infty}^{\infty}
\int_{0}^{\infty} F_m^{\pm}\left( K,\Omega \right)
J_{0}\left( Kr\right) e^{i\Omega t} K dK d\Omega.
\end{equation}
In terms of the vector $F=[F_1^+~F_1^-~F_2^+~F_2^-]^T$
of transformed amplitudes, the linearized system
becomes algebraic and takes the form
\begin{eqnarray}
\left( \begin{array}{cccc}
\Omega_{1}+i\lambda & u_{20}  & u_{10}  & 0 \\
u_{20} & \Omega_{1}-i\lambda & 0 & u_{10} \\
u_{10} & 0  & \Omega_{2}-\delta \Omega +i\lambda & 0\\
 0 & u_{10} & 0 & \Omega_{2}+\delta \Omega-i\lambda
\end{array} \right) F=0,
\nonumber\end{eqnarray}
where $\Omega_mÊ\equiv
\beta_m \Omega^2/2 - \sigma_m K^2/2 - m \beta + (m-1)\delta k$.
The compatibility condition for this $4 \times 4$ system
gives the dispersion relation $\lambda=\lambda(K,\Omega)$.
The eigenvalues $\lambda$
have cumbersome expressions, except for vanishing GVM ($\delta=0$ \cite{noGVM}),
for which we find (set $2f_0=u_{20}^2-2u_{10}^2-\Omega_1-\Omega_2$,
$f_1=(\Omega_1 \Omega_2-u_{10}^2)^2 - u_{20}^2 \Omega_2^2$),
\begin{equation} \label{dispersion}
\lambda=\lambda(K,\Omega)=\pm \sqrt{f_0 \pm \sqrt{f_0^2-f_1} },
\end{equation}
Instabilities sets in when at least one of the four eigenvalues
$\lambda$ in Eq.~(\ref{dispersion}) have a positive real part,
thus causing the perturbation to grow exponentially
with gain $g=g(\Omega,K)=2{\rm Re}(\lambda)$.
Two qualitatively different regimes are found depending on
the sign of GVD. In Fig. 1 we display
the typical spectral gain curve obtained in the
anomalous GVD regime ($\beta_{1,2}=-1$).
As shown, the gain is relatively
narrowband, and its axisymmetric feature
($g$ peaks at $\Omega^2+K^2={\rm constant}$)
reflects the fact that space and time play indeed the same role.
Conversely, when the GVD is normal,
perturbations with large wavevector (angle) $K$ and detuning $\Omega$,
can be amplified. Three different branches of growing
perturbations exist even when the collinear process is
phase-matched ($\delta k=0$), as shown in Fig.2(a).
This type of spatio-temporal MI
entails CCE, which can be regarded as the  amplification process of a superposition
of Bessel $J_0$ beams characterized by progressively larger values of
of angle-frequency ($K-\Omega$) pairs.
The peak gain [obtained in a wide bandwidth of high frequencies $K-\Omega$,
see Fig.2(a)], is displayed in Fig.2(b)
and decreases monotonically for large $|\delta k|$ for both eigenmodes.
\newline\indent
The underlying physics of CCE involve multiple three-photon
processes. Pump photons at $2 \omega$ travelling on-axis decay
into photon pairs at frequency $\omega - \delta \omega$ and
$\omega + \delta \omega$ travelling off-axis with opposite angles.
These in turn can produce off-axis photons at $2\omega \pm \delta
\omega$ via sum-frequency mixing with on-axis pump photons at
$\omega$. In the limit $\delta k=\pm 2$ where the pump degenerate
into the single component $u_{20}^{\pm}=\pm 1$ ($u_{10}^{\pm}=0$),
only the downconversion process takes place, and we can write the
longitudinal projection (the transverse one is identically
satisfied) of the noncollinear phase-matching condition as
$k_z(2\omega)-k_z(\omega + \delta \omega)-k_z(\omega-\delta
\omega)=0$. In the paraxial approximation $k_z(\omega \pm \delta
\omega)=[k(\omega \pm \delta \omega)^2-k_t^2]^{1/2} \simeq
k(\omega \pm \delta \omega)-k_t^2/2k_1$ ($k_t$ is the real-world
tranverse wavenumber) and by expanding $k(\omega \pm \delta
\omega)=k_1 \pm V_{g1}^{-1}\delta \omega+k_1'' \delta \omega^2/2$,
the phase-matching condition yields
\begin{equation}\label{pm}
k_1'' \delta \omega^2 - k_1^{-1}k_t^2= \Delta k.
\end{equation}
In terms of dimensionless frequencies
$K=k_t R_0$ and $\Omega=\delta \omega T_0$,
Eq.~(\ref{pm}) reads as $\Omega^2-K^2=\pm 2$,
which turns out to be in perfect agreement with
the locus $(K-\Omega)$ of peak gain as obtained
from Eq.~(\ref{dispersion}).
\newline\indent
Remarkably we find that CCE is robust against
the presence of GVM. In fact for $\delta \neq
0$ CCE occurs also in the ideal GVD-free limit ($\beta_{1,2}=0$).
The relative magnitude of GVM and GVD depends, however,
on the intensity $I_t$. For instance, in a LBO crystal pumped at $\lambda_0=1~\mu$m
($k_1''=0.045$ ps$^2$/m,  $k_2''=0.25~{\rm ps}^2$/m)
with $I_t=50~{\rm GW/cm}^2$ ($Z_{nl}=0.6$ mm),
we find $Z_w=0.12$ mm which yields $\delta=5.4$.
Figure 3 shows the different gain branches
for this case, when the crystal operates
in the large mismatch limit $k_2-2k_1 = -30~{\rm cm}^{-1}$
($\delta k=-1.8$).  The dimensional gain $G=g/Z_{nl}$
is reported against real-world angles $\theta \simeq
\sin \theta =K/(k_1 R_0)$ and wavelength detunings $\Delta \lambda=
\lambda_0^2 \Omega/(2\pi c T_0)$.
\newline\indent
Though we have considered perturbations growing on top of the SHG
eigenmodes, CCE can occur also from different launching conditions
entailing exchange of energy between $\omega$ and $2\omega$ pump
beams. This case is similar to spatial or temporal MI occurring in
unseeded SHG, and can be analyzed by means of Floquet techniques
\cite{fuerst97,tw97}. In this letter, we rather present numerical
evidence for such phenomenon, by integrating  Eqs.~(1) in 1+1+1D,
setting $\partial_y=0$. As an example Fig.4 shows the
spectral content of the two fields after propagation ($z=15$), as
obtained when only the pump beam $u_1$ is launched at $z=0$, on
top of white noise. As shown, CCE favours the amplification of
spatio-temporal frequencies which closely follow the spectral gain
[see Fig.2(a)]. Due to the large bandwidth of the
process we have also found that CCE takes place from short pulse,
narrow beam excitations, a case which will be discussed elsewhere.
\newline\indent
We end up by emphasizing that CCE in SHG
exhibits important differences with respect to CCE in true Kerr media.
First, SHG makes CCE accessible in materials which are virtually GVD-free,
and, in the large mismatch limit, for both signs
of effective Kerr nonlinearity in the same physical material.
However, the major importance of CCE stems from the fact
that it can trigger the
formation of novel spatio-temporal propagation-invariant wavepackets of Eqs. (1),
i.e. $u_m(r,t,z)=q_m(r,t) \exp(i m \beta z)$ \cite{qels02}.
In the normal GVD regime, $q_m(r,t)$ exist for $\beta<0$ in the form of a
polichromatic superposition of $J_0$ beams \cite{durnin87},
so-called nonlinear X-waves \cite{qels02}, which exist also in the linear limit
\cite{saari97}.  In SHG when the $J_0$ perturbations are amplified at the expense of
an  out-of-phase pump beam, they develop the right sign of nonlinear
phase-shift ($\beta<0$) for the formation of a nonlinear X-wave.
Viceversa, in Kerr media, where nonlinear X-waves also exist,
cw plane-waves and localized wavepackets experience opposite nonlinear
phase-shifts, and such mechanism is unlikely to take place.
\newline\indent
In summary we have shown that spatio-temporal MI of propagation-invariant
as well as dynamically evolving pump beams in SHG gives rise to conical emission,
which might play an important role in the space-time dynamics. Such phenomenon is
intrinsic to quadratic materials and should not be confused with other
recently reported mechanisms of conical emission \cite{moll02}.

Financial support from MIUR (PRIN project), INFM (PAIS project),
UNESCO, MEC (Spain), and Lithuanian Science and Studies Foundation (grant T-491),
is gratefully acknowledged.
The work of C.C. is supported by the Fondazione Tronchetti Provera.


\newpage

Fig. 1. MI gain $g(\Omega,K)$ in the anomalous GVD regime
for the in-phase eigenmode
at phase-matching $\delta k=0$.

Fig. 2. Normal GVD regime ($\beta_{1,2}=1$):
(a) Level plot of conical emission gain $g(\Omega,K)$
for $\phi=0$ eigenmode (similar plot holds for $\phi=\pi$);
(b) Peak value of $g$ vs. $\delta k$.

Fig. 3. Level plot of CCE dimensional gain  $G$ for the
in-phase eigenmode in a LBO crystal (see text for data).

Fig. 4. Spectrally selective amplification of white noise
seen at $z=15$ from unseeded collinearly phase-matched
SHG [$\delta k=0$ in Eqs. (1)] pumped by a cw plane-wave pump at $\omega$: (a)
$|u_1(K,\Omega)|$; (b) $|u_2(K,\Omega)|$.


\begin{thebibliography}{}
\bibitem[\dag]{byline1}
Also with Dept. of Engineering, University of Ferrara.

\bibitem[\ddag]{byline2} Permanent Address:
Dept. Fisica i Enginyeria Nuclear, UPC Terrassa, Spain

\bibitem{tai86} K. Tai, A. Hasegawa, and A. Tomita,
Phys. Rev. Lett. {\bf 56}, 135 (1986).

\bibitem{male01} R. Malendevich, L. Jankovic,
G.I. Stegeman, and J. S. Aitchison,
Opt. Lett. {\bf 26}, 1879 (2001).

\bibitem{trillo95} S. Trillo and P. Ferro,
Opt. Lett. {\bf 20}, 438 (1995).

\bibitem{schiek01}
R. Schiek, H. Fang, R. Malendevich, and G.I. Stegeman,
Phys. Rev. Lett. {\bf 86}, 4528 (2001).

\bibitem{fuerst97} R.A. Fuerst, D.M. Baboiu, B. Lawrence, W.E.
Torruellas, G.I. Stegeman, S. Trillo, and S. Wabnitz,
Phys. Rev. Lett. {\bf 78}, 2756 (1997).

\bibitem{tw97}
S. Trillo, and S. Wabnitz,
Phys. Rev. E {\bf 55}, R4897 (1997).

\bibitem{fang00}
H. Fang, R. Malendevich,
R. Schiek, and G.I. Stegeman,
Opt. Lett. {\bf 25}, 1786 (2000).

\bibitem{liou92} L.W. Liou, X.D. Cao, C.J. McKinstrie,
and G.P. Agrawal, Phys. Rev. A {\bf 46} 4202 (1992).

\bibitem{luther94} G.G. Luther, A.C. Newell, J.V. Moloney,
and E.M. Wright,
Opt. Lett. {\bf 19} 789 (1994).

\bibitem{fonseca99} E.J. Fonseca, B. S. Cavalcanti, J.M. Hickmann,
Opt. Commun. {\bf 169} 199 (1999).

\bibitem{derossi97} A. De Rossi, S. Trillo,
A.V. Buryak, and Yu.S. Kivshar,
Phys. Rev. E {\bf 56}, R4959 (1997).

\bibitem{durnin87} J. Durnin, J.J. Miceli, and J.H. Eberly,
Phys. Rev. Lett. {\bf 58}, 1499 (1987).

\bibitem{noGVM} R.J. Gehr, M.W. Kimmel, and A.V. Smith,
Opt. Lett. {\bf 23}, 1298 (1998).

\bibitem{qels02} C. Conti {\em et al.}
{\it Nonlinear X-waves: light bullets in normally dispersive media ?},
QELS Conference 2002, (Optical Society of America, Washington DC),
paper QTuJ6.

\bibitem{saari97} P. Saari and K. Reivelt,
Phys. Rev. Lett. {\bf 79}, 4135 (1997).

\bibitem{moll02} K.D. Moll, D. Homoelle, A.L. Gaeta,
and R.W. Boyd,
Phys. Rev. Lett. {\bf 88}, 153901 (2002).

\end{thebibliography}
\end{document}